\documentclass[a4paper,aps,pre,superscriptaddress,floatfix,nofootinbib,twocolumn,showpacs,bibtex]{revtex4}

\usepackage[english]{babel}
\usepackage{graphicx}
\usepackage{color}

\begin{document}



\title{Testing the fairness of citation indicators for comparison across scientific domains: the case of fractional citation counts}


\author{Filippo Radicchi}
\affiliation{Howard Hughes Medical Institute (HHMI), Northwestern University, Evanston, Illinois 60208 USA}
\affiliation{Department of Chemical and Biological Engineering, Northwestern University, Evanston, Illinois 60208 USA}
\author{Claudio Castellano}
\affiliation{Istituto dei Sistemi Complessi (ISC-CNR), Via dei Taurini 19, I-00185 Roma, Italy}
\affiliation{Dipartimento di Fisica, ``Sapienza'' Universit\`a di Roma, P.le A. Moro 2, I-00185 Roma, Italy}

\begin{abstract} 
Citation numbers are extensively used for assessing the
quality of scientific research. 
The use of raw citation counts is generally misleading,
especially when applied to cross-disciplinary comparisons, since
the average number of citations received is strongly dependent on the 
scientific discipline of reference of the paper.
Measuring and eliminating biases in citation patterns is crucial
for a fair use of citation numbers.
Several numerical indicators have been introduced with this aim, but
so far a specific statistical test for estimating the fairness 
of these numerical indicators has not been developed.
Here we present a statistical method aimed at estimating the
effectiveness of numerical indicators in the suppression of citation biases.
The method is simple to implement and can be easily generalized to
various scenarios.
As a practical example we test, in a controlled case, the fairness of
fractional citation count, which has been recently proposed as a tool for
cross-discipline comparison.
We show that this indicator is not able to remove biases in citation
patterns and performs much worse than the rescaling of citation counts with
average values.
\end{abstract}

\maketitle

\section{Introduction}
\label{sec:intro}
\noindent
Recent years have witnessed an increasing use of citation numbers
for the quantitative assessment of scientific research activities, and 
many countries have already established national research evaluation
agencies whose judgment criteria are based mostly on citation 
numbers~\citep{gilbert08,abbott09}. 
The use of citation numbers for research evaluation has been 
criticized~\citep{macroberts89,adler09}, especially
because the meaning of citations may be strongly context
dependent~\citep{bornmann08}.
As a matter of fact, however, citations play a crucial role in modern
science, and often important decisions such as
the granting of research funds~\citep{bornmann08a}
or institutional positions~\citep{bornmann06}
are heavily influenced by numerical indicators based on citations.
\\
Generally speaking, citations are interpreted as
proxies for the impact or influence of  papers in the scientific community.
That is,  the more citations a paper has accumulated,
the more relevant this paper can be considered for its own
scientific community of reference. Citation numbers, however,
are not only used for the quantitative evaluation
of scientific publications, but also for the formulation and
quantification 
of numerical indicators devoted at the assessment
of the career of scholars~\citep{hirsch05,egghe06} and
the quality of scientific journals~\citep{garfield06}.
Sometimes the use of citation numbers is also 
extended to the judgment of departments~\citep{davis84,leydesdorff11b}, 
of institutions~\citep{kinney07}
and even of entire countries~\citep{king04}.

\

\noindent
A fundamental problem
in citation analysis is the presence of biases in citation numbers. 
It is for example well known that
papers in mathematics accumulate citations at a rate much lower than
papers in, say, chemistry.
It is therefore unfair to directly compare citation numbers
in mathematics and chemistry. 
Suitably modified indicators should instead be used
in order to remove such patent bias among disciplines.
If such biases are not removed then they can affect comparisons
from the level of individuals up to research groups or institutions.
While a direct comparison among scholars in different disciplines
may seem not so common (although examples exist, see for example
{\tt http://www.topitalianscientists.org/Top\_italian\\
\_scientists\_VIA-Academy.aspx})
when departments, universities or scientific
institutions are compared, as it often occurs, this problem
is unavoidable and potentially very dangerous. 
The lack of a proper handling of this bias may make those comparisons
almost pointless.

\noindent Slightly less severe, the problem of different citation patterns also
exists  for different fields within the same
discipline~\citep{radicchi11}, where it
is customary that citation records of scholars are compared
in competitions for the same resources, such as academic positions or
research grants.
The problem of biases in citation numbers
is therefore crucial, if not in cross-disciplinary
comparisons, in comparisons among
sub-fields or topics of research within the same scientific domain.

\

\noindent
Several studies have dealt with this 
issue~\citep{shubert86,shubert96,vinkler96,vinkler03}.
The common idea is the development of normalized indicators 
(i.e., the raw number of citations is divided
by a discipline dependent factor) able to suppress discrepancies
among scientific domains.
(For other approaches see~\cite{bornmann11,leydesdorff11c}).
Independently of the particular recipe proposed, these studies
do not generally
provide a quantitative test able to determine whether their proposed
method is able to effectively suppress citation biases or not~\footnote{One of the few exceptions to this general trend is the statistical
test performed by~\citet{leydesdorff11a} in order
to check whether their normalization procedure
is able to suppress discipline related biases in a novel formulation
of the journal impact factor.}.
\\
One of the problematic issues for rescaled indicators is the
attribution of papers to disciplines.
This categorization is usually derived from an existing (and questioned)
attribution of journals to disciplines: if a paper appears in a journal
it is assumed to belong to the same category (or categories) the
journal belongs to.
This procedure has several obvious potential inconvenients~\citep{zhou11}.
To overcome this problem, \citet{leydesdorff10c}
have recently proposed  an indicator based on a fractional citation
count~\citep{small85},
i.e., weighting each citation as $1/n$, where $n$ is the total number
of references in the citing paper.
Assuming that differences between citation patterns across domains
are due to different typical lengths in reference lists,
this method provides an implicit normalization of citation counts
that does not require any explicit classification of papers into categories.

\

\noindent In this paper, we contribute to the search for effective ways
to remove the bias in two ways.
\\
First, we propose a general method for testing the effectiveness
of numerical indicators aimed at the removal of
biases in citation counts among scientific domains.
The method relies on a simple selection process and compares the values
of the indicator under test with those expected under the hypothesis
that the indicator is not biased.
Indicators able to suppress citation biases
should produce results statistically consistent
with an unbiased selection process, while their failure
in the test directly indicates their non-effectiveness
in the suppression of biases.
\\
Secondly, as a practical application, we apply the method to two
recently proposed normalization schemes for paper citation counts.
We consider a large database of physics papers, which has the
important feature that the attribution of papers to categories is directly
provided by authors and thus can be considered to be accurate.
In this way the categorization step is not a potential source of problems.
We show that, while the rescaled indicator of ~\citet{radicchi08}
effectively allows an unbiased comparison among different sub-fields,
the fractional citation count of ~\citet{leydesdorff10c} largely
fails the same test and does not constitute a substantial improvement
over raw citation numbers.

\

\noindent
We would like to stress that our notion of 
``fairness'' is based on the strong assumption 
that each discipline or field of research
has the same importance for the development of
scientific knowledge. According
to our assumption, a ``fair'' numerical
indicator based on citation numbers assumes
values that do not depend on the particular scientific domain
taken under consideration. It is clear that our notion
of fairness strongly depend on the
classification of papers into categories (disciplines,
fields, topics). Also it is important to remark that
other possible definitions of fairness could be stated
without relying on the assumption that each discipline
or research field has the same weight for scientific development. 

\section{Material and Methods}
\label{sec:mat}
\noindent
We consider all papers published in journals of the American Physical 
Society (APS, {\tt www.aps.org})
between years $1985$ and $2009$. 
The {\it xml} file containing all the relevant
information about publications in APS journals up to $2009$ was
directly provided by the editorial office of APS 
\\({\tt http://publish.aps.org/datasets-announcement}).
We restrict our analysis only
to standard research publications (Letters, Rapid communications, Brief
Reports and Regular Articles) 
and exclude other type of published material (Editorials, Reviews,
Comments, Replies and Errata)
which may show distinct citation patterns.
The journals considered in our analysis are:
Physical Review Letters,
Physical Review A, Physical Review B,
Physical Review C, Physical Review D  and Physical Review E.
APS journals are the most typical publication outlets
in physics and cover all sub-fields of this discipline.
They therefore represent an optimal benchmark for the study 
of citation patterns of publications within physics~\citep{Redner05}.
The classification of papers into distinct categories is provided by PACS
numbers, which are alphanumerical codes denoting the topic discussed in
the paper and are attributed to papers by authors themselves.
PACS codes are composed of three fields $XX.YY.ZZ$,
where the first two are numerical (two digits each) and the third is
alphanumerical. For our purpose we consider only the first digit of
the $XX$ code, which provides a classification into
very broad categories ({\tt http://publish.aps.org/PACS}).
Hence, for example, two papers with PACS codes
$05.70.Ln$ and $02.50.2r$  both belong to the category $00$, while a 
paper with PACS number $64.60.Ht$ is part of the category $60$.

The first year of the temporal range of our analysis
is $1985$, because in that year PACS 
numbers started to be systematically used.
We consider only papers classified according to the PACS codes,
which are the vast majority 
($>95 \%$ between $1985$ and $2009$, nowadays the selection
by authors of at least one PACS number is compulsory at the submission stage) 
of all papers published in APS journals, for a total of $307,992$
papers. 
In general, authors assign to a paper two or three PACS numbers.
In our analysis, we classify papers only according to their principal
PACS number.
This set of articles represents our set of cited papers,
and here we will study only the properties
of the citations received by these papers.
For this purpose, each article in the set of cited
papers was also retrieved in the
WebOfScience (WOS, {\tt www.isiknowledge.com}) database. 
We collected several information, but mainly 
their unique WOS identification numbers and their total 
number of cites (i.e., the field ``times cited''). 

\

\begin{figure}[!ht]
\includegraphics[width=0.45\textwidth]{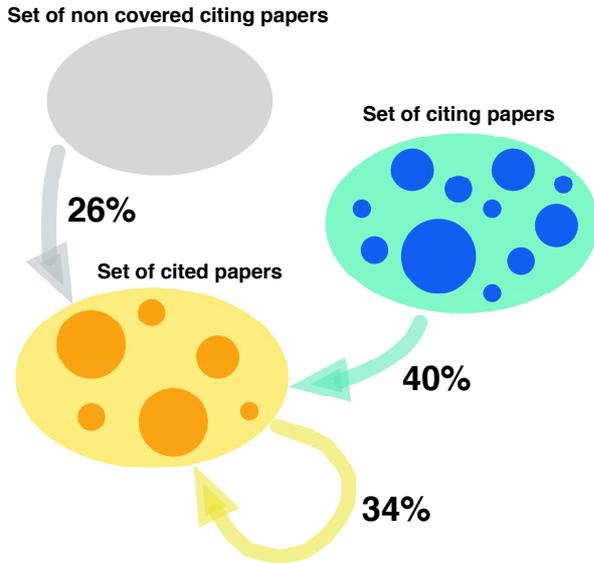}
\caption{Schematic representation of the bibliographic
dataset used in this paper. We focus our attention
on the citation pattern of the set cited papers, which
consists of all papers published in journals of the APS
between years $1985$ and $2009$. Papers in the cited set
are classified in research topics, according
to their principal PACS number. The $34\%$ of the citations
received by the set of cited papers is originated
by papers published between $1985$ and $2011$
in the same journals. We also
consider an additional set of papers, published in
other $139$ scientific journals between $1985$
and $2011$, which cover an additional $40\%$ of
the total citations received by papers in the set
of cited papers. The remaining $26\%$ of the citations
is not included in our analysis.}
\label{fig:ill}
\end{figure}

\noindent As set of citing papers, we
first consider $349,285$ papers published in the same
journals as those belonging to the set of cited papers, but
published between $1985$ and the beginning of $2011$.
For each of these papers we obtain from WOS
their list of references. In the reference list
provided by WOS, cited papers are reported as unique WOS
identification numbers. This information enables
to match referenced papers with those belonging to the set
of cited papers. 
With this first step, we are able to cover about the $34\%$
of the total number of citations declared by WOS, calculated as
the sum of all cites indicated in the field ``times cited''
of the set of cited papers. In order to increase
the coverage of the set of citing papers,
we collect also the list of references of
$1,768,222$ additional articles published in other $139$ journals
between $1985$ and the beginning of $2011$.
These journals are selected because listed as top
citing journals to the set of cited journals in the $2000$ edition
of Journal of Citation Reports (JCR) database.
Please note that this list includes not only physics journals,
but also chemistry, biology, engineering  and multidisciplinary journals.
The inclusion of this new set of citing articles enables us to
cover about the $74\%$ of the total number of citations received
by the set of the cited papers (see Figure~\ref{fig:ill}). 

\

\noindent The entire data collection of
cited and citing papers  was performed between 
March $20$ and March $29$, $2011$. 


%
%

\section{Theory/calculation}
\label{sec:th}
\noindent

\subsection{Normalization methods}
\label{sec:norm}
\noindent 
We consider two normalization procedures which have been
recently proposed for cross-discipline comparisons
of paper citation counts by~\cite{leydesdorff10c} 
and~\cite{radicchi08}, respectively. 
The two schemes are conceptually different:
in the former, citation weights
are functions of the papers from which citations
are originated; in the latter, citation weights
are functions of the papers receiving citations.
In the following, we provide detailed descriptions
of both methods.

\subsubsection{Fractional citation count}
\noindent 
\cite{leydesdorff10c} have recently proposed a
novel normalization scheme aimed at eliminating differences
in citation counts among papers belonging to different scientific disciplines
or about different topics. The method
is based on the intuitive assumption that papers tend to cite
other articles dealing with similar topics.
The whole citation network is therefore organized
into clusters of elements with reasonable high density
of internal citations, while citations among different clusters are more rare. 
Under these assumptions, the average number of citations received
by papers in a given compartment is proportional to the typical
length of the reference list of articles published in that compartment.
Papers in mathematics receive less citations than papers
in biology because a typical paper in mathematics has a shorter
reference list than a typical paper in biology.
The method by~\cite{leydesdorff10c}, called "fractional citation count'',
simply weights a  citation from paper $i$ to paper $j$ 
as $1/n_i$, where $n_i$ is the total number
of articles referenced by paper $i$.
In fractional citation count, the value ${\tilde c}_j$
of the indicator for paper $j$ is given by the
sum of all citations received by paper $j$, where
each citation is opportunely divided
by the number of references of the citing paper
\begin{equation}
\tilde{c}_j = \sum_i \; \frac{A_{ij}}{n_i} \;\;, 
\end{equation}
where $A_{ij}=1$ if paper $i$ cites paper $j$, while $A_{ij}=0$ 
otherwise.
\\
The method has the great advantage that it
does not formally require any {\it a priori} classification
of the papers into scientific domains. This is a great advantage
in many situations, because obtaining a reasonable
classification of papers is very often nontrivial. 
This method does not require any  external information regarding the
classification of papers in different classes, but, as the authors claim,
fractional citation counts ``automatically'' include 
the typical feature of the citation pattern of the cited papers.
The practical disadvantage of the method is however
the necessity of considering the whole list 
of citing papers, which may be difficult to retrieve.
\\
The method has been already applied to the evaluation
of departments in universities~\citep{zhou11},
to a new estimation of the impact factor
of scientific journals~\citep{leydesdorff11a, andres11},
and to a novel formulation of the $h$-index~\citep{andres11}.
Similarly, the indicators developed
by~\citet{zitt10} and~\citet{moed10} for the
assessment of the impact of scientific journals 
are based on a source normalization scheme for citations.
The difference with respect to the indicator
based on fractional citation count 
is, however, that the normalizing factor
is not the exact length of the reference list of the citing
paper, but instead the average length of the reference lists
of all papers published in the same journal as the one
of the citing paper. These studies propose
interesting alternatives to the impact factor but also
stress the inability of fractional citation count to
well account for the degree of cross-field citations
and the growth of the literature in a field of
research.

\subsubsection{Rescaled citation count}
\noindent
A different approach aimed at suppressing
citation biases in cross-disciplinary comparisons 
is the one originally proposed by~\cite{lundberg07}
and then considered also by~\cite{radicchi08}.  
Assuming that papers are classified in compartments
corresponding to scientific
disciplines and fixed years of publication, 
the authors showed that the only relevant
difference in citation patterns corresponding
to different compartments is the value
of a single number $c_0$, the average number of citations
received by papers published in a given scientific
discipline and in a given
year of publication. By assigning to each paper a relative
citation count indicator $c_f=c/c_0$ defined as the total number of cites $c$
received by the paper divided by the value of $c_0$
corresponding to the category and year of publication of the paper,
~\cite{radicchi08} were able to show that this quantity
obeys a probability density function that is universal
among scientific disciplines. The indicator
based on rescaled citation counts thus provides a natural
way of eliminating biases among scientific disciplines, since
the probability to have a paper with relative indicator $c_f$ equal to
a certain value no longer depends on the particular discipline under
consideration.
\\
\noindent
The practical disadvantage of the rescaled citation count
indicator is related to the potential
difficulties that may arise in the classification
of papers in scientific disciplines. In general,
classifications are made at the level of
scientific journals, and this may
lead to some inconsistencies in the classification
of papers. Journals belong to more 
than a scientific discipline if they publish
papers about different subjects. Then, all papers published
in these journals will belong to more
scientific disciplines because their classification
is based on the classification of the journals
where they were published, but in reality
each of these papers is just about a particular
scientific subject and their multi-disciplinary
feature is just an artifact of the classification
procedure.
On the other hand, the results of~\cite{radicchi08} show
a very interesting and not trivial feature of
the citation habits in science:
apart from a typical citation scale, which is discipline
dependent, the way citations accumulate is the
same in all disciplines. More practically,
the universal shape of the citation 
distribution (i.e., a lognormal distribution) allows
to estimate the confidence intervals
of the rescaled citation count indicator~\citep{castellano09}.
\\
\noindent
In the original paper,~\cite{radicchi08} analyzed
$14$ different scientific disciplines. The analysis
has been recently extended to more complete datasets by
~\cite{alba11} and~\cite{waltman11b}, showing
that in general the universality of the citation distribution holds
in many scientific disciplines, with the notable exception of many
social sciences~\citep{waltman11b}. Rescaled
citation counts have been applied also to more
refined contexts for the quantification of scientific
relevance of papers in sub-topics within the same
discipline: \cite{bornmann09} focused their attention on papers
published in chemistry journals, while \cite{radicchi11} on
papers published in journals of physics.

\subsection{Fairness of the indicators}
\label{sec:fairness}
\noindent According to our definition,
the value of a fair indicator associated
to a paper should not depend on the particular category
(topic of research or scientific discipline) of the paper.
In other words, the probability 
of finding a paper with a particular value
of a fair indicator must not depend on the topic/discipline
of research of the paper, it must be the same across fields of research
or scientific disciplines. The ``fairness'' of a citation indicator
is therefore directly quantifiable by
looking at the ability of the indicator
to suppress any potential citation bias
related to the classification of papers in disciplines
or topics of research.
\\
\noindent
Based on these assumptions, here we propose a simple statistical 
test able to assess  the fairness of a citation indicator.
The procedure is very general and can
be simply applied to any type of classification
and/or any type of numerical indicator based
on citations.
\\
\noindent
Imagine to have a set of $N$ total papers
divided in $G$ different categories. Indicate with $N_g$ the number of
papers belonging to the $g$-th category. Each paper in the entire
set has associated a score
calculated according to the rules of the particular
indicator we want to test. 
Imagine now to extract the top $z\%$ of papers
from the whole set of papers. The list
of the top $z\%$ papers in the dataset is composed
of the $n^{(z)}=\lfloor z\, N \, / \, 100 \rfloor$ papers
with the highest values of the score ($\lfloor x \rfloor$ 
stands for the largest 
integer number smaller than or equal to $x$). 
If the numerical indicator is fair,
the presence in the top $z\%$ should not depend on the
particular category to which the paper belongs.
That is, the presence of an article of the $g$-th category
in the top $z\%$ should depend only on the number $N_g$
of papers in category $g$, and not
on the fact that papers in category $g$ may be privileged
for some other particular reason. 
Under these conditions, the number of papers $m^{(z)}_g$ in category 
$g$ that are part of the top $z\%$ of the whole ranking
is a random variate obeying the hypergeometric 
distribution~\footnote{A more general treatment
of the problem would require the use of the multivariate 
hypergeometric distribution
\[
P\left(m^{(z)}_1,  \ldots, m^{(z)}_G\left|n^{(z)}, N, N_1, \ldots, N_G\right.\right) = 
\prod_{g=1}^G {N_g \choose m^{(z)}_g} \left/\; {N \choose n^{(z)}} \right. 
\;\; ,
\]
with $\sum_{g=1}^G m^{(z)}_g = n^{(z)}$ and $\sum_{g=1}^G N_g = N$.
Here however, we perform only independent tests of
fairness and consider one category at time.
}
\begin{equation}
P\left(m^{(z)}_g\left|n^{(z)}, N, N_g\right.\right) = {N_g \choose m^{(z)}_g} \; {N-N_g \choose n^{(z)}-m^{(z)}_g}\; \left/\; {N \choose n^{(z)}} \right. \;\; .
\label{eq:hyper}
\end{equation}
${x \choose y} = x! /\left[y!\, \left(x-y\right)!\right]$ is a
binomial coefficient which calculates the total number of
ways in which $y$ elements can be extracted out $x$ total elements.
Eq.~(\ref{eq:hyper}) describes a simple urn model~\citep{polya}, where
elements (i.e., papers in our case) are randomly extracted
from the urn without replacement. With this statistical
model we can simply calculate the expected number of
papers in category $g$ present in the top $z\%$ as $E\left(m^{(z)}_g\right)=
n^{(z)}\,N_g/N$. Moreover, we can make use
of Eq.~(\ref{eq:hyper}) for estimating confidence intervals
or other relevant statistical quantities.

\section{Results}
\label{sec:res}
\noindent 
We base our entire analysis on the bibliographic data set
described in sec.~\ref{sec:mat}. A category
here is therefore intended as the collection of all papers published
in the same year and with the same first digit of the principal PACS number.
This means that we have at our disposal
a total of $250$ different categories:
$25$ different years of publication and $10$ different PACS numbers.
We also stress once more that the analysis is based
on the $76\%$ of the total number of citations
effectively accumulated by the papers in our set of cited papers.
With our data, we are in fact not able
to identify the source for the remaining $24\%$ of the citations.
This problem affects only the computation of the numerical indicator
based on fractional   citation counts, but for consistency we prefer to use
the same amount of information also for the 
computation of raw and rescaled citation counts.

\

\begin{figure}[!ht]
\includegraphics[width=0.45\textwidth]{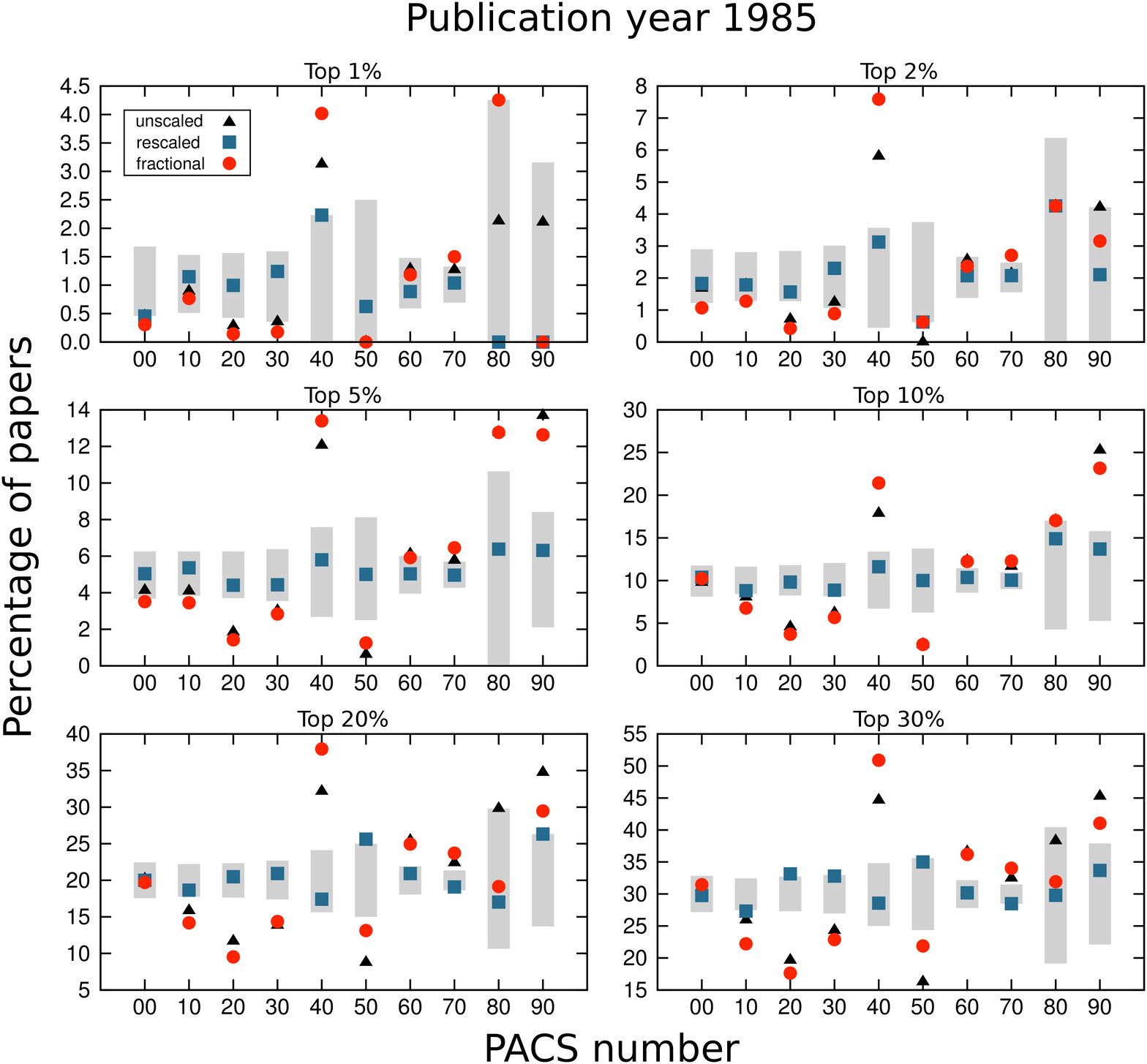}
\caption{Percentage of papers belonging to the
top $z\%$ for different PACS classification codes and different
values of $z$. Here we consider only papers published in 
year $1985$. Black triangles represent
the results obtained with raw citation counts,
blue squares stand for the results 
obtained with the indicator based
of rescaled citation counts, while
red circles indicate the results obtained using the
indicator based on fractional citation counts.
Gray areas bound the $90\%$ confidence intervals
and are calculated using  Eq.~(\ref{eq:hyper}).
The values of
the average number of citations
$c_0$ and the number of papers
$N$ considered for each PACS code are: 
PACS $00$ $c_0=32.73$ and $N=655$,
PACS $10$ $c_0=29.48$ and $N=783$,
PACS $20$ $c_0=21.30$ and $N=703$,
PACS $30$ $c_0=25.71$ and $N=564$,
PACS $40$ $c_0=58.75$ and $N=224$,
PACS $50$ $c_0=18.13$ and $N=160$,
PACS $60$ $c_0=39.48$ and $N=1,014$,
PACS $70$ $c_0=37.99$ and $N=1,734$,
PACS $80$ $c_0=46.04$ and $N=47$,
PACS $90$ $c_0=57.00$ and $N=95$.
}
\label{fig1}
\end{figure}

\begin{figure}[!htb]
\includegraphics[width=0.45\textwidth]{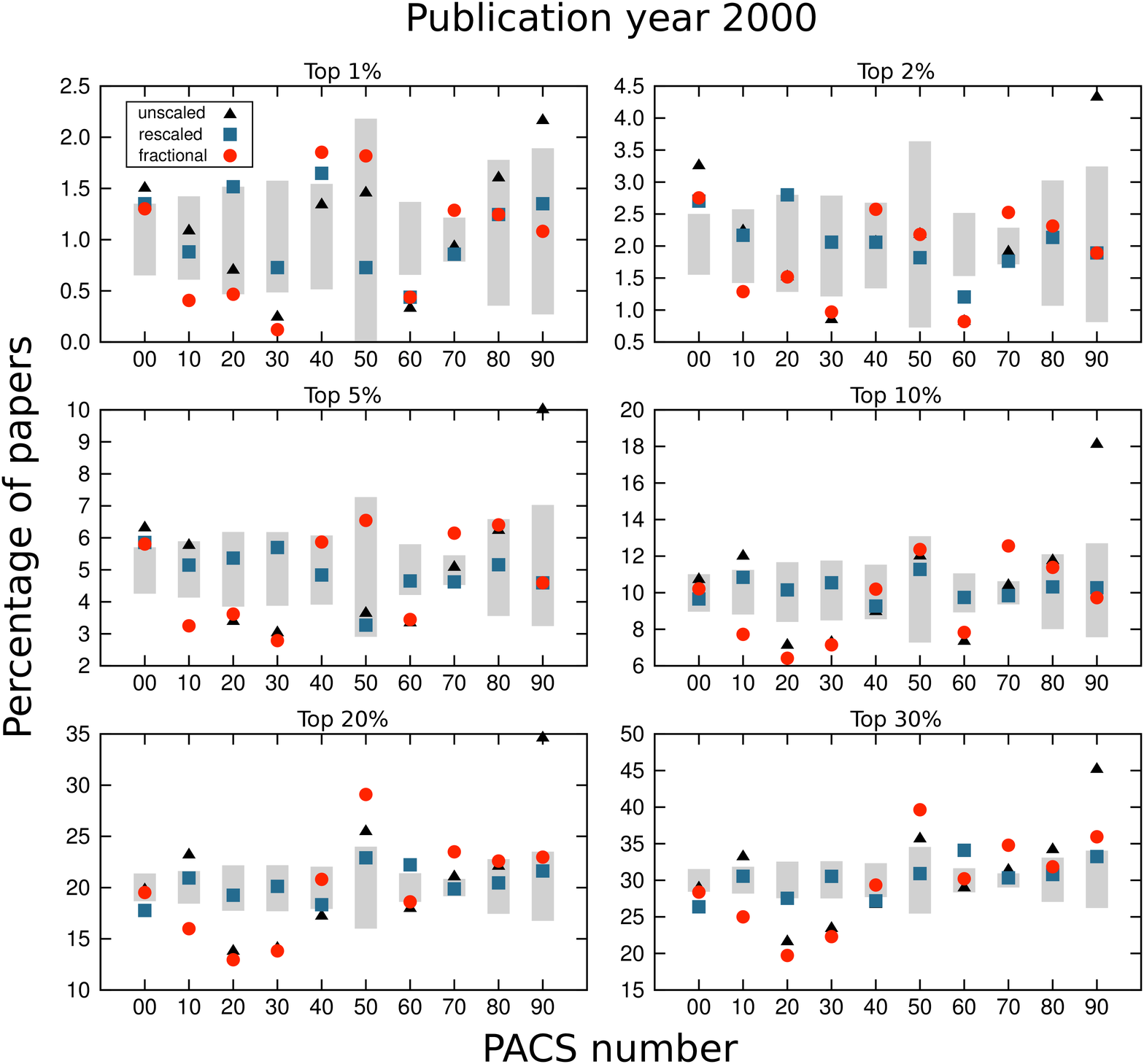}
\caption{Same as Fig.~\ref{fig1}, but for papers published
in year $2000$. The values of
the average number of citations
$c_0$ and the number of papers
$N$ considered for each PACS code are: 
PACS $00$ $c_0=26.65$ and $N=1,998$,
PACS $10$ $c_0=26.48$ and $N=1,476$,
PACS $20$ $c_0=19.43$ and $N=857$,
PACS $30$ $c_0=19.08$ and $N=825$,
PACS $40$ $c_0=24.02$ and $N=971$,
PACS $50$ $c_0=27.29$ and $N=275$,
PACS $60$ $c_0=21.59$ and $N=1,827$,
PACS $70$ $c_0=25.92$ and $N=4,197$,
PACS $80$ $c_0=26.75$ and $N=562$,
PACS $90$ $c_0=38.10$ and $N=370$.
}
\label{fig2}
\end{figure}

\noindent
In Figures~\ref{fig1}
and~\ref{fig2}, we report the 
percentage of papers associated to a particular PACS number
that are present in the top $z\%$ of all papers published in
a given year. We show the results only for papers published in
years $1985$ (Fig.~\ref{fig1}) and $2000$ (Fig.~\ref{fig2}),
but qualitatively similar results are obtained
when we consider other publication years. 
In general, we see that the use of
raw citation numbers causes clear disproportions
among different subjects of research. Papers in 
``nuclear physics'' (PACS $20$) are underrepresented
in the top percentage of papers, because papers in this
sub-field of physics are typically less cited than
papers in other sub-fields. On the other hand,
papers in ``astronomy \& astrophysics'' (PACS $40$)
over-populate the set of highly cited papers. The proportion
of papers belonging to this subject of research are typically
two to three times larger than what expected on average
in the case of an unbiased selection process. At the same
time, the indicator based on fractional citation count still leads 
to ``unfair'' results.
Some PACS numbers (the same
as those privileged by raw citation counts) are favored, 
and the percentage of papers in these categories belonging to
the top $z\%$ is much higher
than what can be predicted (higher than the value
corresponding to the $95\%$ confidence interval).
Conversely, other PACS numbers are underrepresented
and their percentage is lower than the $5\%$
confidence bound. 
The indicator based
on rescaled citations, on the other
hand, works pretty well. The results obtained are
in the majority of the cases compatible with 
the statistical model of Eq.~(\ref{eq:hyper}). 
The result holds for almost all PACS numbers
and does not depend on the number of papers belonging
to the PACS.
\\
\begin{figure}[!htb]
\includegraphics[width=0.45\textwidth]{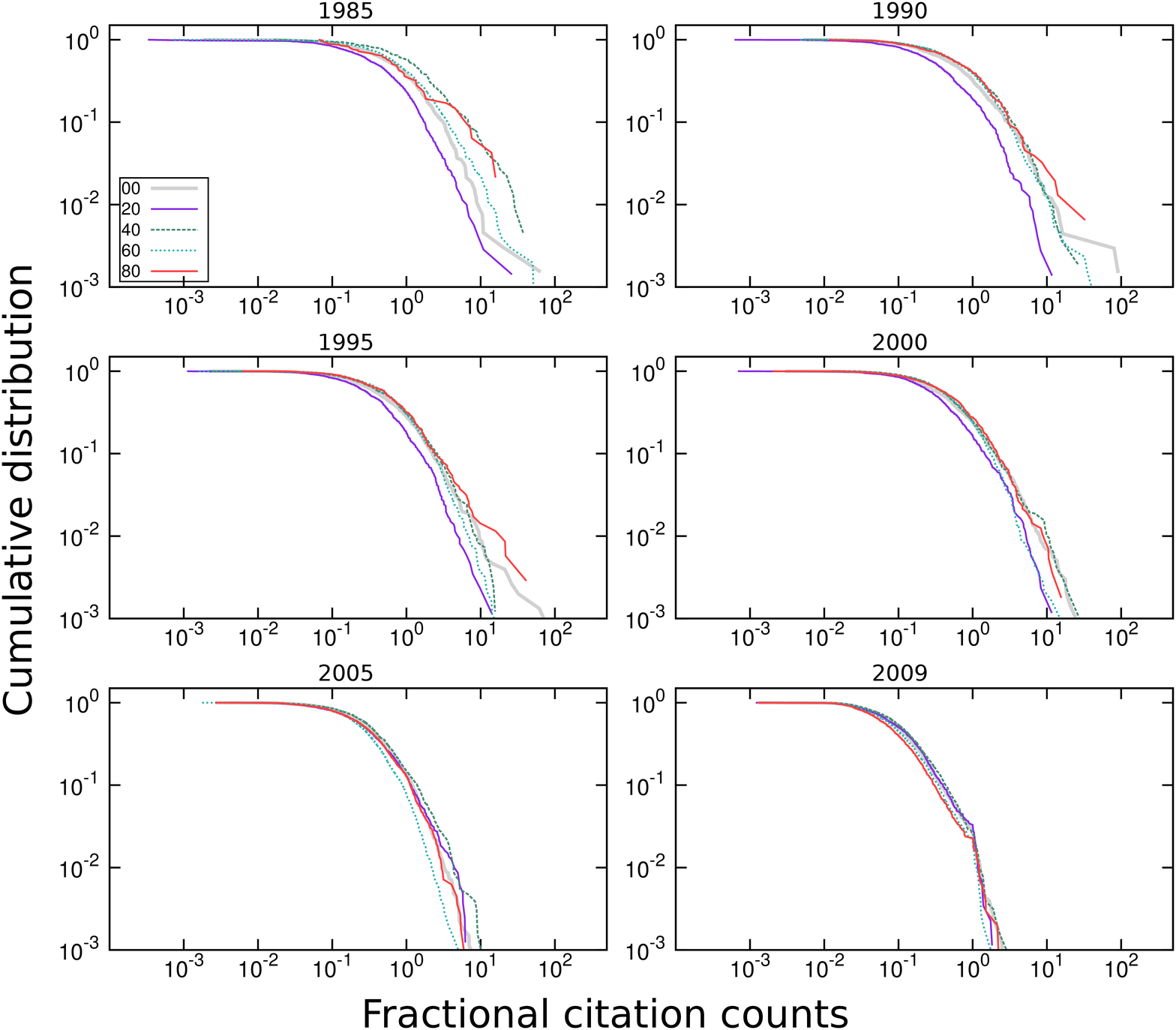}
\caption{Cumulative distribution of fractional citation counts
 for different PACS numbers and year of publication.}
\label{fig3}
\end{figure}

\begin{figure}[!htb]
\includegraphics[width=0.45\textwidth]{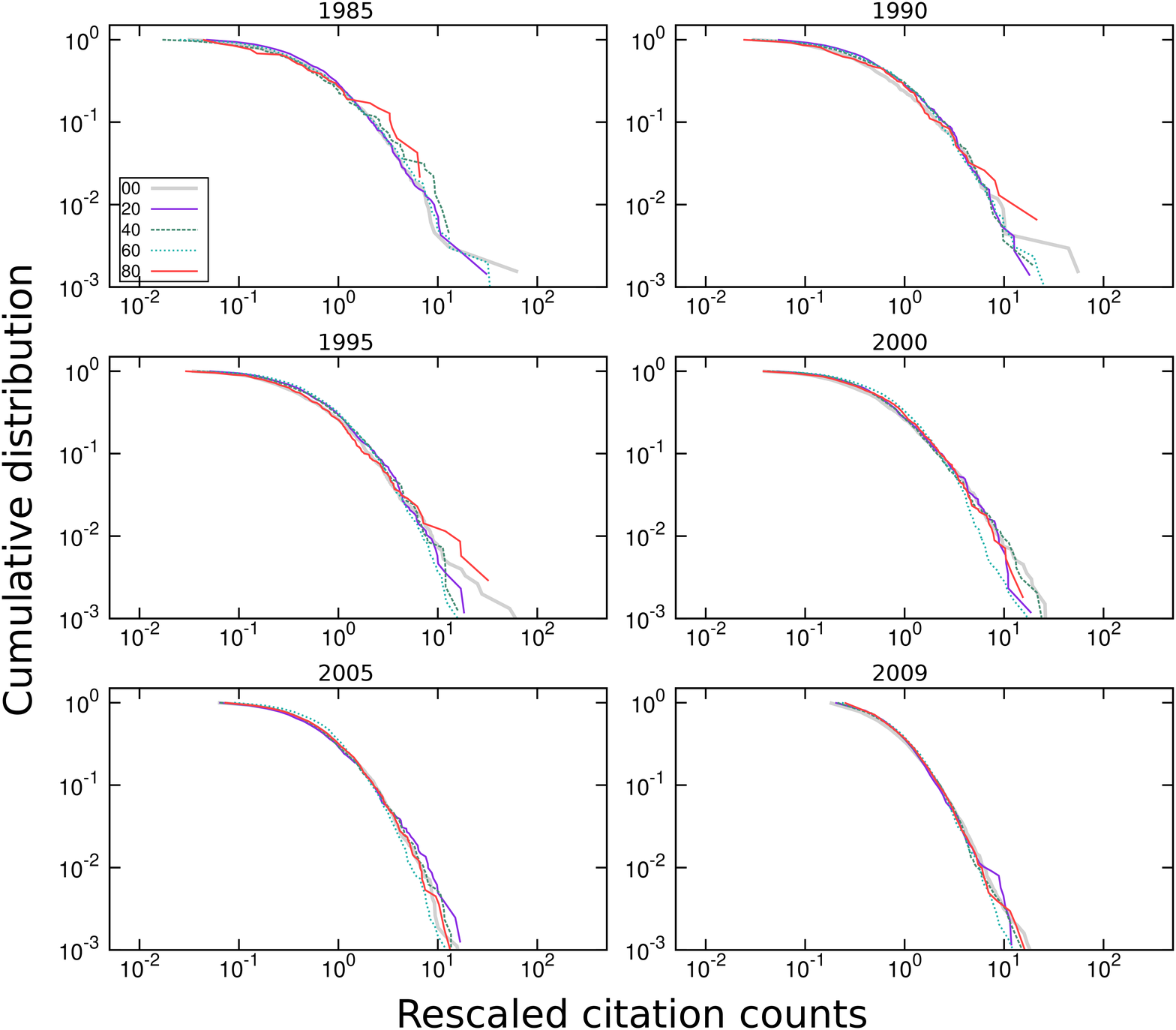}
\caption{Cumulative distribution of the rescaled number of citations
for different PACS numbers and year of publication.}
\label{fig4}
\end{figure}

The same results can be understood in a more
intuitive manner by looking at Figures~\ref{fig3}
and~\ref{fig4}. The cumulative distributions, 
relative to different PACS numbers,
of the indicator based on fractional citation count 
do not collapse on top of each other.
There is in general a systematic bias, and papers
published under a particular PACS number have associated
larger values of the indicator. The presence
of a bias is particularly
evident in the top left panel of Fig.~\ref{fig3}.
Here, we consider only papers published in $1985$
and report the cumulative distribution for PACS
numbers $00$, $20$, $40$, $60$ and $80$. The indicator
based of fractional citation counts is constantly larger for
papers published under PACS
$40$, followed by those belonging to PACS $80$, $60$,
$00$ and $20$.
\\
The same systematic shift is not visible
in the equivalent plots obtained with
rescaled citations (Fig.~\ref{fig4}).
In this case, the curves corresponding to different
PACS numbers overlap in a consistent way:
Rescaled citations do not favor any particular field.
\\
A quantitative measure summarizing the global performance of
the different indicators is presented in Table~\ref{table}.
We calculate, for all $250$ sets under consideration (identified by
the principal PACS number and the publication year),
the percentage of categories for which the fraction of papers in
the top $z\%$ falls within the $90\%$ confidence interval of the
distribution in the hypothesis that the indicator is not biased.
It turns out that the rescaled citation count fully removes the
bias for values of $z$ up to $10\%$, while fractional citation count
perform much worse (and only marginally better than raw citations).

\begin{table}
\begin{center}
\begin{tabular}{c c c c}
z & Rescaled citations & Fractional citations & Raw citations \\ \hline
\hline
$1$  & $88\%$ & $61\%$ & $60\%$ \\
\hline
$2$  & $90\%$ & $56\%$ & $50\%$ \\
\hline
$5$  & $92\%$ & $49\%$ & $41\%$ \\
\hline
$10$ & $92\%$ & $48\%$ & $37\%$ \\
\hline
$20$ & $79\%$ & $40\%$ & $33\%$ \\
\hline
$30$ & $67\%$ & $34\%$ & $30\%$  \\ \hline
\hline
\end{tabular}
\end{center}
\caption{Fraction of all categories for which the number of papers belonging
to the top $z\%$ falls within the $90\%$ confidence interval denoted by
the gray areas in Figs.~\ref{fig1} 
and~\ref{fig2}.}
\label{table}
\end{table}

\section{Discussion and conclusions}
\label{sec:dis}
\noindent
The results presented in the previous section show that
attributing to each citation a fractional weight equal
to the inverse of the total number of references is not
sufficient to remove the biases that
make citation numbers large in some disciplines (or
fields) and small in others.
This conclusion has been obtained by defining
a quantitative statistical procedure to test the fairness
of generic numerical indicators for the impact of papers
and applying it in a controlled case. Our notion 
of fairness is based on the assumption
that each scientific discipline contributes equally 
to the development of scientific knowledge,
and therefore a fair numerical indicator based on citation counts
should assume values that do not depend on the particular discipline
taken under consideration. In this sense, 
fractional citation counts bring only a modest improvement
with respect to the use of raw citation numbers.
On the other hand it turns out that the rescaling of the
number of cites with average values of the reference set
is remarkably successful in removing biases. This indicator
should then be used for a fair comparison of the impact of
papers across disciplines.

\

\begin{figure}[!htb]
\includegraphics[width=0.45\textwidth]{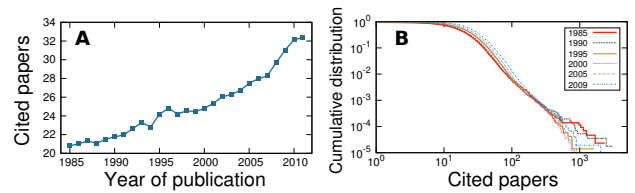}
\caption{{\bf(A)} Average number of cited papers as a function
of the publication year of the citing article. {\bf(B)}
Cumulative distribution of the number of cited papers
for different years of publication. Both figures are based
on the set of citing articles, and numbers refer to
their entire reference lists.}
\label{fig5}
\end{figure}

\noindent Counting citations fractionally is not an effective way
to remove biases. 
Nevertheless, it would be very beneficial from
a different point of view.
As shown in Fig.~\ref{fig5}A, which refer to the entire data set of
citing articles,
papers published in $1985$ cited on average $21$ other publications, while
for papers  published in $2011$ the average length of the reference list
exceeds $32$. 
Even more striking is the shape of the distribution
of the number of papers cited by a single publication
(see Fig.~\ref{fig5}B).
The length of the reference list is very broadly distributed,
with a nonegligible probability to observe
papers citing thousands of documents.
In practice this means that a single publication can contribute
to citation numbers more than a hundred of others together.
The large variability
in the length of the reference lists is
due to the heterogeneity of the type of
citing documents. Short communications or letters, for example,
are often subjected to length restrictions and 
therefore can cite only a limited amount
of other papers. On the other hand, review articles have much longer
lists of references which sometimes can be even two order 
of magnitude larger
than those of shorter documents. Nevertheless, the length
of reference lists is growing with time, 
and some form of citing-side normalization would discourage the
inflation of reference lists, thus making citation counts 
(in a different sense) more fair.



\end{document}